\documentclass[12pt,letterpaper,english]{article}
\usepackage[T1]{fontenc}

\usepackage{float} %for graphs
\usepackage{verbatim}
\usepackage{booktabs}
\usepackage{graphicx}
\usepackage{titlesec}
\usepackage[margin=1.in]{geometry}
\usepackage{bbold}
\usepackage{amsmath, amssymb, mathrsfs}
\usepackage{natbib}
\usepackage{babel}
\usepackage{setspace}\doublespace%\setstretch{2.4}
\usepackage[labelsep=period,labelfont=bf]{caption}
\usepackage{subcaption}
\usepackage{epstopdf}
\usepackage{titlesec}

\providecommand{\U}[1]{\protect\rule{.1in}{.1in}}
\def \b{\textbf}
\def \tx{\textrm}

\DeclareGraphicsExtensions{.pdf,.png,.jpg}
\titleformat{\section}[block]{\Large\bfseries\filcenter}{\thesection.\quad}{1em}{}

\begin{document}

\title{A Bayesian Non-Parametric Approach to Asymmetric Dynamic Conditional Correlation Model With Application to Portfolio Selection}

\author{
  Audron\.{e} Virbickait\.{e}\\ 
\small{Universidad Carlos III de Madrid}\\ \small{ Getafe (Madrid), Spain, 28903}\\ \small{audrone.virbickaite@uc3m.es}  
 \\ \and M.\ Concepci\'on Aus\'in\\\small{Universidad Carlos III de Madrid}\\ \small{Getafe (Madrid), Spain, 28903}\\\small{ concepcion.ausin@uc3m.es} 
  \and Pedro Galeano\\\small{Universidad Carlos III de Madrid}\\ \small{Getafe (Madrid), Spain, 28903}\\ \small{pedro.galeano@uc3m.es}  
}

\date{}

\maketitle
\thispagestyle{empty}
\newpage

\begin{abstract}

We propose a Bayesian non-parametric approach for modeling the distribution of multiple returns. In particular, we use an asymmetric dynamic conditional correlation (ADCC) model to estimate the time-varying correlations of financial returns where the individual volatilities are driven by GJR-GARCH models. The ADCC-GJR-GARCH model takes into consideration the asymmetries in individual assets' volatilities, as well as in the correlations. The errors are modeled using a Dirichlet location-scale mixture of multivariate Gaussian distributions allowing for a great flexibility in the return distribution in terms of skewness and kurtosis. Model estimation and prediction are developed using MCMC methods based on slice sampling techniques. We carry out a simulation study to illustrate the flexibility of the proposed approach.  We find that the proposed DPM model is able to adapt to several frequently used distribution models and also accurately estimates the posterior distribution of the volatilities of the returns, without assuming any underlying distribution. Finally, we present a financial application using Apple and NASDAQ Industrial index data to solve a portfolio allocation problem.  We find that imposing a restrictive parametric distribution can result into underestimation of the portfolio variance, whereas DPM model is able to overcome this problem.

\noindent \textbf{Keywords:} Bayesian Analysis; Dirichlet Process Mixtures; Markov Chain Monte Carlo; Multivariate GARCH; Portfolio Allocation.

\noindent \textbf{JEL Classification:} C11, C32, C53, C58, G11.

\end{abstract}
\thispagestyle{empty}
\newpage

\section{INTRODUCTION}\label{S:Introduction}

Modeling the stylized features of the assets' returns has been extensively researched for decades and the topic yet remains of great interest. The ARCH-family models, first introduced by \cite{Engle1982} and then generalized by \cite{Bollerslev1986}, are without doubt the most analyzed and used in practice to explain time-varying volatilities, see \cite{Bollerslev1992}, \cite{Bollerslev1994}, \cite{Engle2002b}, \cite{Terasvirta2009} and \cite{Tsay2010}, for instance. When dealing with multiple returns, one must also take into consideration the mutual dependence between them, see \cite{Bauwens2006}, \cite{Silvennoinen2009} and \cite{Tsay2010}, for instance. In particular, conditional correlation models firstly proposed by \cite{Engle2002}, \cite{Tse2002} and \cite{Christodoulakis2002}, play an important role because there is evidence that conditional correlations are time dependent. The asymmetric behavior of individual returns has been well established in the financial literature, see \cite{Hentschel1995}. More recently, \cite{Cappiello2006} have pointed out that conditional correlations may exhibit some stylized
features, such as persistence and asymmetry and, consequently, have proposed Asymmetric Dynamic Conditional Correlation (ADCC) models for time-varying correlations.

It is well known, that every prediction, in order to be useful, has to come with a certain precision measurement. In this way the agent can know the uncertainty of the risk she is facing. In the field of MGARCH models, the distribution of the returns, that strongly depends on the distributional assumptions for the error term, permits to quantify this uncertainty about the future. However, the traditional premises of multivariate Normal or Student-t distributions may be rather restrictive because usually the empirical distribution of returns is slightly skewed and their tails are fatter than those of a Gaussian distribution, see \cite{Rossi2010}, for example. Alternative parametric choices such as
the  Student-t density, see \cite{Fiorentini2003}, the  skew-Student-t distribution, see \cite{Bauwens2005}, or finite mixtures of  Gaussian distributions, see e.g.\ \cite{Ausin2007} and \cite{Galeano2010}, have been proposed in the literature that usually improve the fit of the GARCH models. However, all of them require the assumption of a certain parametric model.

In this paper, in order to model the stylized features of the assets' returns, we assume a model which allows for asymmetries not only in individual assets' returns, but also in their correlations. In particular, we consider an Asymmetric Dynamic Conditional Correlation (ADCC) model for time-varying correlations, proposed by \cite{Cappiello2006}, with GJR-GARCH models, proposed by \cite{Glosten1993}, for individual volatilities. This specification, denoted by ADCC-GJR-GARCH, provides a much more realistic evaluation of the co-movements of the assets' returns than standard symmetric MGARCH models. Additionally, we propose a Bayesian non-parametric approach for the
ADCC-GJR-GARCH model avoiding the specification of a particular parametric distribution for the return innovations. More specifically, we consider a Dirichlet Process Mixture (DPM) model of Gaussian distributions, firstly
introduced by \cite{Antoniak1974}. This is a very flexible model that can be viewed as an infinite location-scale mixture of Gaussian distributions which includes, among others, the Gaussian, Student-t, logistic, double
exponential, Cauchy and generalized hyperbolic distributions, among others. We follow closely the works of \cite{Ausin2014}, who have applied the DPM models for univariate GJR-GARCH, and \cite{Jensen2013}, who have used DPM models
for the multivariate symmetric DVEC by \cite{Ding2001}. 
Non-parametric  time-varying volatility models have been of great interest in the recent literature, not only in GARCH, but also in Stochastic Volatility setting, see \cite{Jensen2010}, \cite{Jensen2012}, \cite{Delatola2011} and \cite{Delatola2013}.

The Bayesian approach also helps to deal with parameter uncertainty in portfolio decision problems, see e.g.\ \cite{Jorion1986}, \cite{Greyserman2006}, \cite{Avramov2010} and \cite{Kang2011}, among others. This is in contrast with the usual maximum likelihood estimation approach, which assumes a "certainty equivalence'' viewpoint, where the sample estimates are treated as the true values, which is not always correct and has been criticized in a number of papers. As noted by \cite{Jorion1986}, this estimation error can gravely distort optimal portfolio selection. In this paper, we propose a Bayesian method which provides the posterior distributions of the one-step-ahead optimal portfolio weights, which are more informative than simple point estimates. In particular, using the proposed approach, it is possible to obtain Bayesian credible intervals for the optimal portfolio weights. Note that
the Bayesian methodology also provides some other advantages over the classical maximum likelihood techniques, see \cite{Ardia2010}. For example, it is easy to incorporate via priors complicated positivity constraints on the parameters to ensure positive variance and covariance stationarity. Additionally, it is possible to approximate the posterior distribution of any other non-linear function of the parameters, as will be done for the optimal portfolio weights. For a survey on Bayesian inference methods for univariate and multivariate GARCH models see \cite{Virbickaite2013}.

Therefore, the main contribution of this work is the proposal of a Bayesian nonparametric method for explaining the dynamics of the assets' returns via an ADCC-GJR-GARCH model with the use of DPM location-scale mixture models for the return innovations. Also, we present an application of Bayesian non-parametric techniques in portfolio decision problems and explore the differences in uncertainty between the proposed approach and conventional restrictive distributional assumptions, where the objective is to provide a more realistic evaluation of risk of financial decisions. As commented before, this study extends the work by \cite{Ausin2014} to the multivariate framework and the recent work by \cite{Jensen2013} to the asymmetric setting. Also, differently from the work of \cite{Jensen2013}, we always assume a conjugate prior specification and we use a different sampling approach.

The outline of the paper is as follows:  Section \ref{S:Model_inf_pred} describes the model, inference and prediction from a Bayesian perspective. Section \ref{S:Applications} introduces the time-varying portfolio optimization problem. Section \ref{S:Simulation} presents a short simulation study. Section \ref{S:data_results} illustrates the proposed approach using a real data example, solves a portfolio allocation problem and carries out model comparison. Finally, Section \ref{S:Conclusion} concludes.

\section{MODEL, INFERENCE AND PREDICTION}\label{S:Model_inf_pred}

This section describes the asymmetric dynamic conditional correlation GJR-GARCH model used for volatilities and correlations. Then, we introduce the DPM specification for the error term. Finally, we provide a detailed explanation of the implementation of Bayesian non-parametric inference and the methodology of obtaining predictive densities of returns and volatilities.

\subsection{The ADCC-GJR-GARCH-DPM Model}\label{ss:asymmetric_dcc}

As commented before, financial returns usually exhibit two types of asymmetries: in individual volatilities and in conditional correlations. Therefore, in one hand, we choose the GJR-GARCH model proposed by \cite{Glosten1993} for individual returns, to incorporate asymmetric volatility effects, while, on the other hand, we use the ADCC model proposed by \cite{Cappiello2006} and based on the previous work by \cite{Engle2002}, to model joint volatilities. Then, we assume that the vector of $K$ asset returns is given by:
\begin{equation}
r_{t}=H_{t}^{1/2}\epsilon _{t},  \label{F:gjr}
\end{equation}
for $t=1,\ldots,T$, where $H_{t}$ is a scale symmetric $K\times K$ matrix and $\epsilon _{t}$ are a sequence of iid random variables with an unknown $K$-dimensional distribution $\mathcal{F}_{K}$, for which we will assume a DPM
prior specified later. As usual in all the DCC models, the matrix $H_{t}$ can be decomposed as follows:
\begin{equation}
H_{t}=D_{t}R_{t}D_{t},  \label{Ht}
\end{equation}
where $D_{t}$ is a diagonal matrix that contains the square root of the elements of the main diagonal of $H_{t}$, denoted by $d_{iit}$, for $i=1,\ldots ,K$. These elements are assumed to follow GJR-GARCH models given by:
\begin{equation}
d_{iit}^{2}=\omega _{i}+(\alpha _{i}+\phi _{i}I_{i,t-1})r_{it-1}^{2}+\beta_{i}d_{iit-1}^{2},  \label{dii2}
\end{equation}
with parameters $\omega _{i},\alpha _{i},\phi _{i},\beta _{i}>0$, for $i=1,\ldots ,K$ and where $I_{i,t-1}$ is an indicator function such that $I_{i,t-1}=1$ if $r_{it}<0$ and is $I_{i,t-1}=0$, otherwise. On the other hand, to introduce $R_{t}$, we need to define
\begin{equation}
\varepsilon _{t}=D_{t}^{-1}r_{t},\text{ and }\eta _{t}=\varepsilon _{t}\odot
I(\varepsilon _{t}<0),  \label{eps}
\end{equation}%
where $\odot $ denotes Hadamard matrix product operator and $I(\varepsilon_{t}<0)$ is a vector with $i$th component equal to 1 if $\varepsilon _{it}<0$, and $0$ otherwise. Then, $R_{t}$ is given by:
\begin{equation}
R_{t}=Q_{t}^{\star -1}Q_{t}Q_{t}^{\star -1},  \label{model9}
\end{equation}
where $Q_{t}$ is the $K\times K$ matrix given by:
\begin{equation}
Q_{t}=S(1-\kappa -\lambda -\delta /2)+\kappa \times \varepsilon_{t-1}\varepsilon _{t-1}^{\prime }+\lambda \times Q_{t-1}+\delta \times \eta_{t-1}\eta _{t-1}^{\prime},  \label{Qt}
\end{equation}
where $S$ is a sample correlation matrix of $\varepsilon _{t}$, and $Q_{t}^{\star}$ is a diagonal matrix with the square root of the $i$th diagonal element of $Q_{t}$ on its $i$th diagonal position. We impose that $\kappa ,\lambda
,\delta >0$ and $\kappa +\lambda +\delta /2<1$ to ensure the positivity and stationarity of $Q_{t}$. Finally, the vector $\Phi =\left( \omega ,\alpha,\beta ,\phi ,\kappa ,\lambda ,\delta \right) $ summarizes the set of
parameters describing the matrices $H_{t}$, for $t=1,2,\ldots$

As for the unknown distribution of $\epsilon_t\sim \mathcal{F}_K$, there has been and ongoing discussing about the best specification for the heavy-tailed  financial returns. Next, we present a flexible DPM specification for the errors and some of the most important special cases arising from this model. Using the stick-breaking representation by \cite{Sethuraman1994}, a DPM of Gaussian distributions can be expressed as a location-scale Gaussian mixture model with infinitely many components and therefore, it can be easily defined as an extension of a parametric mixture model. The base distribution of the DP, usually denoted by $G_0$, corresponds to the prior distribution of the component parameters in the infinite mixture. The concentration parameter, denoted by $c$, can be interpreted as the prior belief on the number of clusters in the mixture. Small values of $c$ assumes a priori an infinite mixture model with a small number of components with large weights. On the contrary, large values of $c$ assumes a priori an infinite mixture model with all the weights being very small.

Therefore, the resulting density function of $\epsilon_t$ can be written as:
\begin{equation}
f\left(\epsilon_t|\rho,\mu,\Lambda\right) =  \sum\limits_{j=1}^{\infty}\rho_j \mathcal{N}_K\left(\epsilon_t|\mu_j,\Lambda_j^{-1}\right),\label{f(e_t)}
\end{equation}
where $\mathcal{N}_K$ denotes a $K$-variate normal density. Let us denote by $\Omega = \left\{{\rho_j,\mu_j,\Lambda_j}\right\}_{j=1}^{\infty}$  the infinite-dimensional parameter vector describing the innovation mixture distribution. Here $\rho_j$, represent the component weights, $\mu_j$ are the component means and $\Lambda_j$ are the precision matrices, for $j=1,2,\ldots$ Using the stick breaking representation, the weights of the infinite mixture components are reparameterized as follows: $\rho_1=v_1$, $\rho_j=(1-v_1)\hdots(1-v_{j-1})v_j$, where a Beta prior distribution is assumed for $v_j\sim \mathcal{B}(1,c)$, for ${j=1,2,\hdots}$ Clearly, there will be some sensitivity to the choice of the concentration parameter $c$. Therefore, we further assume a Gamma hyper-prior distribution for $c$, $c\sim \mathcal{G}(a_0,b_0)$. Finally, as a base distribution, we assume a conjugate Normal-Wishart prior for $(\mu _{j},\Lambda _{j}) \sim \mathcal{NW}(m _{0},s_{0},W_0,d_0)$, where:
\begin{align*}
\mu _{j}|\Lambda_j & \sim \mathcal{N}_{K}\left( m
_{0},(s_{0}\Lambda_j )^{-1}\right) , \\
\Lambda _{j}& \sim \mathcal{W}(W_0,d_0),
\end{align*}%
for $j=1,2,\hdots$, such that $\tx{E}\left[\Lambda_j\right] = d_0\times W_0^{-1}$ and $\tx{E}\left[\Lambda_j^{-1}\right] = \left(d_0-(K+1)/2\right)^{-1}\times W_0$.

In summary, the complete set of model parameters is denoted by $\Theta = (\Phi,\Omega)$. Given the information available up to time $t-1$, denoted by $r^{t-1}$, the conditional density of the returns can be written as follows:
\begin{equation}\label{F:density_returns}
f(r_t|\Theta,r^{t-1})=\sum\limits_{j=1}^{\infty}\rho_j \mathcal{N}_K\left(r_t|H_t^{1/2}\mu_j, H_t^{1/2}\Lambda_j^{-1}(H_t^{1/2})' \right),
\end{equation}
with conditional mean given by:
\begin{equation}\label{admean}
\mu_{t}^{\star}=E\left[ r_{t}|\Theta,r^{t-1}\right] =H_{t}^{1/2}\sum_{j=1}^{\infty }\rho _{j}\mu_{j}
\end{equation}
and conditional covariance matrix:
\begin{equation}\label{advolat}
H_{t}^{\star}=\tx{Cov}\left[ r_{t}|\Theta,r^{t-1}\right] =H_{t}^{1/2}\tx{Cov}\left[ \epsilon_{t}|\Omega\right] (H_{t}^{1/2})^{\prime },
\end{equation}
where
\begin{equation*}
\tx{Cov}\left[ \epsilon _{t}|\Omega \right] =\sum_{j=1}^{\infty }\rho _{j}\left( \Lambda_{j}^{-1}+\mu _{j}(\mu _{j})^{\prime }\right) -\left( \sum_{j=1}^{\infty}\rho _{j}\mu _{j}\right) \left( \sum_{j=1}^{\infty }\rho
_{j}\mu_{j}\right) ^{\prime }.
\end{equation*}

It is important to notice that this full unrestricted model induces GARCH-in-Mean effects, since the conditional mean of the returns is not restricted to be zero. Moreover, the DPM model for $\epsilon _{t}$ does not assume
an identity covariance matrix. As noted in \cite{Jensen2013}, imposing moment restrictions in DPM models is still an open question. However, the prior information considered essentially centers $\epsilon _{t}$ around an identity covariance matrix.

On the other hand, an essential issue in choosing more complicated models versus the simple ones is the ability to handle numerous assets. The DPM model is very flexible in this sense, since the general specification described before contains numerous other simplified models. For example, it clearly contains the single Gaussian as a special case when the first mixture weight is equal to one. Also, it is possible to impose a symmetric distribution for the innovations by simply assuming that the mixture means are all equal and, in particular, it could be reasonable to impose $\mu_j=0$, for $j=1,2,\ldots$. If we further assume that the precision matrices are all diagonal, $\Lambda_j = \tx{diag} \left(\lambda_{j1}, \hdots,\lambda_{jK}\right)$,  this will lead to uncorrelated innovations. Finally, we could in addition assume that the diagonal elements of each precision matrix are all equal by considering $\Lambda_j = \lambda_jI_K$. In this paper we will use the full version of the DPM model to illustrate the flexibility of it. However, the adaptation of  the model to these particular cases in order to simplify the problem of many assets is straightforward.

\subsection{MCMC algorithm}\label{ss:bayesian}

The following section describes a Markov Chain Monte Carlo (MCMC) algorithm to sample from the posterior distribution of the parameters of the ADCC-GJR-GARCH-DPM model introduced in the previous section.  The algorithm is based on the procedure by \cite{Walker2007}, who introduces slice sampling schemes to deal with the infiniteness in DPM, the retrospective MCMC method of \cite{Papaspiliopoulos2008} and the ideas by \cite{Papaspiliopoulos} who combines these two methods to obtain a new composite algorithm, which is better, faster and easier to implement. Generally, all these approaches compared to traditional schemes based on the original algorithm by \cite{Escobar1995} produce better mixing and simpler algorithms.

Following \cite{Walker2007}, in order not to sample an infinite number of values at each MCMC step, we introduce a latent variable $u_t$, such that the joint density of $(\epsilon_t, u_t)$ given $\Omega$ is:
\begin{equation}\label{F:joint_eps_u}
f(\epsilon_t,u_t|\Omega)=\sum_{j=1}^{\infty}\b{1}(u_t<\rho_j)\mathcal{N}_K(\epsilon_t|\mu_j,\Lambda_j^{-1}).
\end{equation}
Let $A_{\rho}(u_t)=\{j:\rho_j>u_t\}$ be a set of size $N_{u_t}$, which is finite for all $u_t>0$. Then the joint density of $(\epsilon_t,u_t)$ in \eqref{F:joint_eps_u} can be equivalently written as $f(\epsilon_t,u_t|\Omega)=\sum_{j\in A_{\rho}(u_t)}\mathcal{N}_K(\epsilon_t|\mu_j,\Lambda_j^{-1})$. Integrating over $u_t$ gives us the  density of infinite mixture of distributions (\ref{f(e_t)}). Finally, given $u_t$, the number of mixture components is finite. In order to simplify the likelihood, we also need to introduce a further indicator latent variable $z_t$, which indicates the mixture component that $\epsilon_t$ comes from: $f(\epsilon_t,z_t=j,u_t|\Omega)=\mathcal{N}_K(\epsilon_t|\mu,\Lambda^{-1})\b{1}(j\in A_{\rho}(u_t))$. Then, the log-likelihood of $\Theta$, given the latent variables $u_t$ and $z_t$ looks as follows:
\begin{equation}\label{likelihood}
\log L(\Theta|\left\{{r_t,u_t,z_t}\right\}_{t=1}^T)=-\frac{1}{2}\sum\limits_{t=1}^T \left(K\log(2\pi)+\log|H_{t,z_t}^{\star}|+(r_t-\mu_{t,z_t}^{\star}) H_{t,z_t}^{{\star}-1} (r_t-\mu_{t,z_t}^{\star})'\right),
\end{equation}
where $\mu_{t,z_t}^{\star}$ and $H_{t,z_t}^{\star}$ are the conditional mean vector and conditional covariance matrix given $z_t$, i.e.:
\begin{gather*}\label{equation:adjusted_mean}
\mu_{t,z_t}^{\star} =   H_t^{1/2} \mu_{z_t},\\
H_{t,z_t}^{\star}=H_t^{1/2}\Lambda_{z_t}^{-1}H_t^{1/2}.
\end{gather*}
respectively.

Using these latent variables, we now construct the following MCMC algorithm that is described step by step.

Firstly, given $z_t$, for $t=1,\ldots,T$, the conditional posterior distribution of the concentration parameter, $c$, is independent of the rest of the parameters, as seen in \cite{Escobar1995}. So, we first sample an auxiliary  variable $\xi\sim \mathcal{B}(c+1,T)$ and then $c$ from a Gamma mixture:
\begin{align*}
\pi_{\xi}\mathcal{G}(a_0+z^{\star},b_0-\log(\xi))+(1-\pi_{\xi})\mathcal{G}(a_0+z^{\star}-1,b_0-\log(\xi)),
\end{align*}
where $z^{\star}=\max(z_1,\hdots,z_T)$ and $\pi_{\xi}=(a_0+z^{\star}-1)/(a_0+z^{\star}-1+T(b_0-\log(\xi)))$.

In the second step, we sample from the conditional posterior of $v_j$ for $j=1,\hdots,z^{\star}$, which is given by:
\begin{align*}
v_j|\left\{{z_t}\right\}_{t=1}^T\sim\mathcal{B}(n_j+1,T-\sum\limits_{l=1}^j n_l+c),
\end{align*}
where $n_j$ is the number of observations in the $j$th component and $\sum_{l=1}^j n_l$ gives the cumulative sum of the groups. Also, $\rho_1=v_1$, $\rho_j=(1-v_1)\hdots(1-v_{j-1})v_j$, for ${j=2,\hdots,z^{\star}}$.

In the third step, we sample the uniform latent variables $u_t\sim\mathcal{U}(0,\rho_{z_t})$, for $t=1,\hdots,T$.

In the fourth step, following \cite{Walker2007}, we need to find the smallest $j^{\star}$ such that $\sum_{j=1}^{j^{\star}}\rho_j>u^{\star}$, where $u^{\star}=\min(u_1,\hdots,u_T)$. Then, if $z^{\star}<j^{\star}$, we need to sample $v_j$, for $j=z^{\star}+1,\hdots,j^{\star}$, from the prior and sample $\rho_j$ accordingly.

Next, in the fifth step, we sample from the conditional posterior distribution of the mixture parameters, which are also Normal-Wishart distributions, $(\mu _{j},\Lambda _{j}) \sim \mathcal{NW}(m _{j},s_{j},W_j,d_j)$, for $j=1,\hdots,j^{\star}$, where:
\begin{align*}
m_j &= \frac{s_0 m_0+n_j \bar\epsilon_j}{s_0+n_j} , \qquad s_j = s_0 +n_j,\\
W_j &= W_0^{-1}+S_j+\frac{s_0 n_j}{s_0+n_j}(m_0-\bar\epsilon_j)(m_0-\bar\epsilon_j)',\\
S_j &=\frac{1}{n_j} \sum\limits_{t:z_t=j}^{T}(\epsilon_t - \bar{\epsilon}_j)(\epsilon_t - \bar{\epsilon}_j)',\qquad \bar{\epsilon}_j =\frac{1}{n_j} \sum\limits_{t:z_t=j}^{T}\epsilon_t,\\
d_j&=d_0+n_j.
\end{align*}
Note that this approach is different from the one described in \cite{Jensen2013} since they assume independent prior distributions for $\mu_j$ and $\Lambda_j$, and then, it is necessary to include some Gibbs steps to sample from the conditional posterior.

In the sixth step, we generate to which component each observation belongs to by sampling from the following conditional posterior distribution:
\begin{align*}
\tx{Pr}(z_t=j|...)\propto \b{1}\left(j\in A_{\rho}\left(u_t\right)\right)\mathcal{N}_K(\epsilon_t|\mu_j,\Lambda_j^{-1}),
\end{align*} see also \cite{Walker2007}.

The rest of the steps of the algorithm concern updating the parameters of the ADCC-GJR-GARCH model. For that, we use the Random Walk Metropolis Hasting (RWMH), following a similar procedure as in \cite{Jensen2013}. For the set of parameters $\Phi$ a candidate value $\tilde{\Phi}$ is generated from a $D$-variate Normal distribution with mean equal to the previous value of the parameter, where $D=4K+3$, is the number of parameters in $\Phi$ as follows:
\begin{align*}
\tilde{\Phi}\sim \left\{
\begin{array}{c c c}
  \mathcal{N}_D \left(\Phi,V\right) & \tx{w.p.} & p   \\
  \mathcal{N}_D \left(\Phi,100 V\right) & \tx{w.p.} & 1-p
    \end{array}\right.
\end{align*}
The probability of accepting a proposed value $\tilde{\Phi}$, given the current value $\Phi$, is $\alpha(\Phi,\tilde{\Phi})=\min\left\{1,\prod_{t=1}^T l(r_t|\Phi)/ \prod_{t=1}^T \tilde{l}(r_t|\tilde{\Phi})\right\}$, where the likelihood used is as in \eqref{likelihood}, see e.g.\ \cite{Robert2004}. The covariance matrix $V$ is obtained by running some initial MCMC iterations and then adjusting the sample covariance matrix by some factor in order to achieve the desired acceptance probability. In this paper the acceptance probabilities are adjusted to be between $20\%$ and $50\%$ and we use $p = 0.9$.

\subsection{Prediction}\label{ss:predictive}

In this section, we are mainly interested in estimating the one-step-ahead predictive density of the returns:
\begin{equation}\label{f:predictive}
f(r_{T+1}|r^T)=\int f(r_{T+1}|\Theta,r^T)f(\Theta |r^T)d\Theta,
\end{equation}
where $f(r_{T+1}|\Theta,r^T)$ is specified in \eqref{F:density_returns}. Although this integral is not analytically tractable, it can be in principle approximated using the MCMC output,\begin{equation}\label{F:pred_density}
f(r_{T+1}|r^T)\simeq \frac{1}{M}\sum \limits_{m=1}^M f(r_{T+1}|\Theta^{(m)},r^T),
\end{equation}where $M$ is the length of the MCMC chain and $\Theta^{(m)}$ is the infinite set of parameters at the $m$-th iteration. However, in practice, at each iteration, there are a finite number of weigths, $\rho_j^{(m)}$, for $j=1,\ldots,j^{{\star}(m)}$, and the corresponding pairs of means, $\mu_j^{(m)}$, and precision matrices, $\Lambda_j^{(m)}$. Then, as considered in \cite{Jensen2013}, we can use the following simulation procedure at each MCMC iteration.

Repeat for $r=1,\ldots,R$:
\begin{itemize}
\item [i.] Sample a random variable $a\sim\mathcal{U}(0,1)$.
\item [ii.] Take such $\rho_{r}^{(m)}$ for which $\sum_{j=1}^{r-1}\rho_{j-1}^{(m)}<a<\sum_{j=1}^{r}\rho_j^{(m)}$ and the corresponding $(\mu_r,\Lambda_r)^{(m)}$.
\item [iii.] If $\sum_{j=1}^{j^{{\star}(m)}} \rho_j^{(m)}<a$, sample $\left(\mu_r, \Lambda_r\right)^{(m)}$ from the Normal-Wishart prior.
\end{itemize}

And then, approximate the one-step-ahead density in (\ref{F:pred_density}) by,
\begin{equation}
f(r_{T+1}|\Theta^{(m)},r^T)\simeq \frac{1}{R}\sum \limits_{r=1}^R \mathcal{N}_{K} \left(r_{T+1}|\mu_r^{(m)} H_{T+1}^{(m)1/2},H_{T+1}^{(m)1/2}\left(\Lambda_r^{(m)}\right)^{-1} (H_{T+1}^{(m)1/2})'\right),
\end{equation}
where $\left(\mu_r,\Lambda_r\right)^{(m)}$, are the $R$ pairs of means and precision matrices simulated for $r=1,\ldots,R$, and $H_{T+1}^{(m)}$ is the value of the $H_{T+1}$ matrix at the $m$-th MCMC iteration.

Using this simulation procedure, we can also obtain predictions for many other important measures. For example, the posterior expected value of the adjusted one-step-ahead mean and volatility matrix, introduced in (\ref{admean}) and (\ref{advolat}), respectively, can be approximated by:
\begin{equation}
\tx{E}\left[\mu_{T+1}^{\star}\mid r^{T}\right] \simeq\frac{1}{M}\sum\limits_{m=1}^{M}\mu_{T+1}^{\star(m)},\label{Eadmean}
\end{equation}
and
\begin{equation}
\tx{E}\left[H_{T+1}^{\star}\mid r^{T}\right]\simeq\frac{1}{M}\sum\limits_{m=1}^{M}H_{T+1}^{{\star}(m)},\label{Eadvolat}
\end{equation}
where,

\begin{equation}\label{F:adjusted_mean}
\mu_{T+1}^{\star(m)}=H_{T+1}^{(m)1/2}\left(\frac{1}{R}\sum \limits_{r=1}^R\mu_r^{(m)}\right)
\end{equation}
and
\begin{equation}\label{F:adjusted_variance}
 H_{T+1}^{\star(m)}=H_{T+1}^{\left( m\right) 1/2}\left( \frac{1}{R}\sum_{r=1}^{R}\left( \left(
\Lambda _{r}^{\left( m\right) }\right) ^{-1}+\mu _{r}^{\left( m\right) }(\mu
_{r}^{\left( m\right) })^{\prime }\right) -\left( \frac{1}{R}%
\sum_{r=1}^{R}\mu _{r}^{\left( m\right) }\right) \left( \frac{1}{R}%
\sum_{r=1}^{R}\mu _{r}^{\left( m\right) }\right) ^{\prime }\right) \left(
H_{T+1}^{\left( m\right) 1/2}\right) ^{\prime }.
\end{equation}

In order to obtain the posterior distributions of the adjusted means and volatilities, one should fix a certain  $R$: a number of components  to sample at each iteration $m$. Since the number of components in the data is not known a priori, one might chose $R$ depending on the  number of clusters in the data. However, this implies that there is no upper limit for $R$, which might  result into sampling a very large number of components at each step and increasing computational cost. Instead, we propose to make use of the truncation introduced in \eqref{F:joint_eps_u} and use only $j^{\star}$ components with their corresponding weights $\rho$ at each step, such that equations \eqref{F:adjusted_mean} and \eqref{F:adjusted_variance} become the following:

\begin{equation}
\mu_{T+1}^{\star(m)}=H_{T+1}^{(m)1/2}\left(\sum \limits_{j=1}^{j^{\star(m)}}\rho_i^{(m)}\mu_i^{(m)}\right)
\end{equation}
and
\begin{align}
 &H_{T+1}^{\star(m)}=H_{T+1}^{(m) 1/2}\times\\  \notag
& \left(\sum_{j=1}^{j^{\star (m)}}\rho_i^{(m)}\left( \left(
\Lambda _{i}^{\left( m\right) }\right) ^{-1}+\mu _{i}^{(m)}(\mu
_{i}^{(m)})^{\prime }\right) - 
\left(
\sum_{j=1}^{j^{\star (m)}}\rho_i^{(m)}\mu _{i}^{(m)}\right) \left( 
\sum_{j=1}^{j^{\star (m)}}\rho_i^{(m)}\mu _{i}^{(m)}\right) ^{\prime }\right) 
\left(H_{T+1}^{(m) 1/2}\right) ^{\prime }.
\end{align}

Similarly, we can approximate the posterior median and credible intervals using the quantiles of the posterior samples $\left\{{\mu_{T+1}^{\star(m)}}\right\}_{m=1}^M$ and $\left\{{H_{T+1}^{{\star}(m)} }\right\}_{m=1}^M$.
\section{PORTFOLIO DECISIONS}\label{S:Applications}

As commented in the introduction, optimal asset allocation is greatly affected by the parameter uncertainty, which has been recognized in a number of papers, see \cite{Jorion1986} and \cite{Greyserman2006}, among others. They conclude that in the frequentist setting the estimated parameter values are considered to be the true ones, therefore, the optimal portfolio weights tend to inherit this estimation error. Instead of solving the optimization
problem on the basis of the choice of unique parameter values, the investor can choose the Bayesian approach, because it accounts for parameter uncertainty, as seen in \cite{Kang2011} and \cite{Jacquier2013}, for example.

The main objective of diversification is to reduce investor's exposure to risk. See \cite{Markowitz1952} and  \cite{Merton1972} for some classical portfolio optimization references. Nowadays, there is a wide variety of portfolio optimization objectives, such as maximizing agent's utility or minimizing expected shortfall, among many others. In this paper we consider one of the mostly used objectives, where the investor minimizes the portfolio variance. The Global Minimum Variance (GMV) portfolio can be found at the very peak of the efficient frontier. Given the time series of returns $r_1,\ldots,T$, the standard approach is to consider the unconditional covariance matrix of the returns, $\Sigma=\tx{Cov}[r_t]$ and solve the following optimization problem:
\begin{equation*}
p^{\star}=\tx{arg}\min\limits_{p:p'1_K=1} \tx{Var}[r_t^P],
\end{equation*}
where $p$ is the weight vector, $1_K$ is a K-vector of ones and $r_{T+1}^P=p'r_{T+1}$ is the portfolio return at time point $T+1$. The problem has solution:
\begin{equation*}
p^{\star}=\frac{\Sigma^{-1}1_K}{1'_K\Sigma^{-1}1_K},
\end{equation*}
that is independent of the time point $T$. Note that, if we choose to impose the short sale constraint, i.e., $p_i\geq 0$, $\forall i=1,\hdots,K$, the problem cannot be solved analytically anymore and it requires numerical optimization techniques.

However, recent results suggest that the use of the time-varying covariance matrix to determine portfolio weights leads to better performing portfolios than the use of a constant covariance matrix. For instance, \cite{Giamouridis2007} find that portfolios, constructed under a dynamic approach, have lower average risk and higher out-of-sample risk-adjusted realized return, see also \cite{Yilmaz2011}. \cite{Cecchetti1988} was the first to suggest the use of MGARCH models in optimal allocation context. Since then, there has been a number of papers investigating the differences in estimation and evaluating their performance using various approaches, from simple OLS, to bivariate vector autorregression (VAR), to GARCH. In particular, \cite{Kroner1993}, \cite{Rossi2002} and \cite{Yang2004}, among others, have shown that GARCH-type models leads to the overall portfolio risk reduction.

Consequentely, to solve the portfolio allocation problem in our case, instead of $\Sigma$, we use the adjusted one-step-ahead conditional covariance matrix for the assets returns $H_{T+1}^{{\star}}$, defined in (\ref{advolat}), which varies continuously on the basis of available information up to time $T$, $r^T$. Therefore, we are able to obtain optimal portfolio weights for time point $T+1$ as follows:
\begin{equation}\label{F:gmv}
p_{T+1}^{\star}=\frac{H_{T+1}^{{\star}-1}1_K}{1'_K H_{T+1}^{{\star}-1}1_K}.
\end{equation}

Using the MCMC output, we can obtain samples from the entire posterior distribution of optimal portfolio weights for $T+1$, $f(p_{T+1}^{\star}|r^T)$. This approach relies on solving the allocation problem at every MCMC iteration and approximate for example the posterior mean of the optimal portfolio weights by:
\begin{equation*}
\tx{E}[p_{T+1}^{\star}|r^T]=\int p_{T+1}^{\star} f(\Theta|r^T)d\Theta\approx\frac{1}{M}\sum\limits_{m=1}^{M}p_{T+1}^{{\star}(m)},
\end{equation*}
where $\left\{p_{T+1}^{{\star}(m)}\right\}_{m=1}^M$ is a posterior sample of optimal portfolio weights obtained from (\ref{F:gmv}) for each value of one-step-ahead conditional covariance matrix of the returns, $\left\{H_{T+1}^{{\star}(m)}\right\}_{m=1}^M$, in the MCMC sample. In other words, since we have assembled $M$ one-step-ahead volatility matrices, we can solve the portfolio allocation problem $M$ times. As in the previous section, we can similarly approximate the posterior median of $p_{T+1}^{\star}$ and credible intervals by using the quantiles of the sample of optimal portfolio weights. In this manner, we are able to approximate the  posterior distribution of the optimal portfolio variance, $\sigma_{T+1}^{2{\star}}$, and optimal portfolio gain, $g_{T+1}^{{\star}}$, using the samples $ \left\{(p_{T+1}^{{\star}'} H_{T+1}^{{\star}}p_{T+1}^{\star})^{(m)}\right\}_{m=1}^M$ and $\left\{(p_{T+1}^{{\star}} \mu_{T+1}^{{\star}'})^{(m)}\right\}_{m=1}^M$, respectively, where
\begin{align*}
p\left(\sigma_{t+1,P}|r^{t}\right)&\sim\left\{(\sigma_{t+1,P})^{(m)}\right\}_{m=1}^M = \left\{(p_{t+1}^{\star'} H_{t+1}^{\star}p_{t+1}^{\star})^{(m)}\right\}_{m=1}^M,\\
p\left(r_{t+1,P}|r^t\right)&\sim \left\{(r_{t+1,P})^{(m)}\right\}_{m=1}^M = \left\{(p_{t+1}^{\star} r_{t+1}^{'})^{(m)}\right\}_{m=1}^M.
\end{align*}

\section{SIMULATION STUDY}\label{S:Simulation}

The goal of this simulation study is to show the flexibility and adaptability of the DPM specification for the innovations for the ADCC-GJR-GARCH model introduced Section \ref{S:Model_inf_pred}. For this, we consider three bivariate time series of $3000$ observations simulated from a ADCC-GJR-GARCH model with the following innovation distributions: (a) Gaussian $\mathcal{N}(0,I_2)$; (b) Student-t $\mathcal{T}(I_2,\nu=8)$; (c) Mixture of two bivariate Normals $0.9\mathcal{N}(0,\sigma^2_{1}=0.8,\sigma_{12}=0.0849,\sigma^2_{2}=0.9)+0.1\mathcal{N}(0,\sigma^2_{1}=2.8,\sigma_{12}=-0.7637,\sigma^2_{2}=1.9)$. Notice that these are most frequently used distributional assumptions for the financial return data. 

Note that, in the third case, we have chosen larger variances for the second mixture component to allow for the presence of extreme returns but preserving an identity covariance matrix. Then, we estimate all three data sets using the proposed ADCC-GJR-GARCH-DPM model assuming uninformative uniform priors restricted to the stationary region for $\Phi$ and setting $m_0 = 0_2$, $s_0 = 0.1$, $d_0 = 5$, $W_0 = I_2/5$, $a_0 = 4$ and $b_0 = 4$. The MCMC algorithm is run for 10,000 burn-in plus 40,000 iterations. The point estimates are not reported in the paper to save space. All parameters were estimated well, with true parameters always inside the $95\%$ credible intervals.

\begin{figure}[H]
\begin{center}
\caption{Contour plots of the true and estimated one-step-ahead predictive densities, $f(r_{T+1}\mid r^T)$, for the three simulated data sets.}
\label{figure:illustration_contours}
\includegraphics[width=1\textwidth]{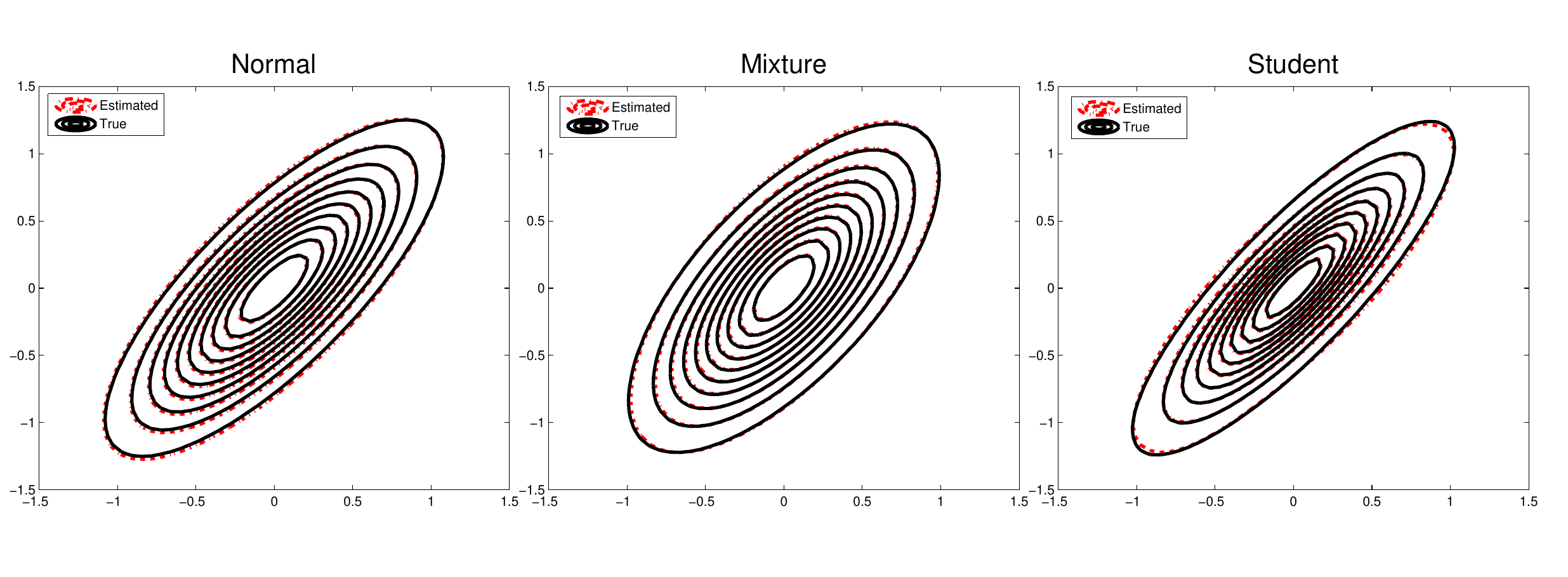}
\end{center}
\end{figure}

Figure \ref{figure:illustration_contours} presents the contour plots that compare the true one-step-ahead predictive densities of returns, given the model parameters, with the estimated ones, obtained from (\ref{F:pred_density}) by setting $R=3$. As we can see, the estimated contours of the one-step ahead return densities are very close to the true ones. Note that these contours can be seen as a summary of the estimation results for all 11 parameters of the model
$\Phi=\left(\omega,\alpha,\beta,\phi,\kappa,\lambda,\delta\right)$ and the distribution for the error term. Therefore, it seems that the proposed infinite mixture model is a very flexible tool that is able to adjust to rather
different return specifications. This is of primary interest because in practice one never knows which is the true error distribution. Therefore, the proposed approach appears to be able to fit adequately several frequently used
distributions.

First part of Table \ref{table:illustration_clusters} presents the estimated posterior mean, median and $95\%$ credible intervals for the number of clusters, $z^{\star}$, for the three generated datasets. For the Gaussian dataset, the proposed DPM model estimates very few non-empty components, $1.23$ on average, where there is always a clear dominant weight. For the Student-t dataset, the proposed DPM model estimates a large number of clusters, around $19.66$,
with similar small weights. This is expected since, as commented in \cite{Jensen2013}, the Student-t distribution can be viewed as a limiting case of a DPM model when the concentration parameter goes to infinity and, consequently, the number of clusters increases indefinitely. Finally, for the two-component mixture data, the DPM model can identify very well the two underlying clusters with posterior mean around $2.68$. The second and third
part of Table \ref{table:illustration_clusters} also show the estimation results for the concentration parameter, $c$, and its transformed value $A=c/(1+c)$, where $0<A<1$,  that has been used by \cite{Jensen2013} to provide an intuition of the probability of having infinite clusters in the mixture. However, note that, different to \cite{Jensen2013}, we have previously defined a Gamma prior on $c$ instead of a uniform prior on $A$. Observe that the obtained results are coherent, since the posterior means of $c$ and $A$ for the Gaussian case are the smaller ones ($A = 0.2509$), respectively, while for the Student-t case are the largest ones ($A=0.6892$), respectively. Finally, for the two-component mixture dataset, the
posterior means of $c$ and $A$ are between the corresponding values of the Gaussian and Student-t cases, that can be seen as a compromise between the two extreme cases.

\begin{table}[tbp]
\caption{Posterior mean, median and $95\%$ credible intervals for the number of non-empty clusters, $z^{\star}$, concentration parameter, $c$, and quantity $A$, for the three simulated data sets.}
\centering
\label{table:illustration_clusters}
\begin{footnotesize}
\begin{tabular}{c c c c c c c}
\hline\hline
  \toprule
  & \multicolumn{2}{c}{Gaussian}& \multicolumn{2}{c}{Student-t} & \multicolumn{2}{c}{2 comp. mixture}  \\
&Mean & $95\%$ CI &Mean & $95\%$ CI&Mean & $95\%$ CI\\
&Median& &Median &&Median &
\\\hline
$z^{\star}$ & 1.2330 & (1.0000, 3.0000) & 19.6612 & (9.0000, 33.0000) & 2.6765 & (2.0000, 5.0000) \\
& 1.0000 & & 19.0000 & & 2.0000 &\\
$c$& 0.3578 & (0.0934, 0.8072) & 2.5037 & (0.9327, 5.0059) & 0.4863 & (0.1512, 1.0408)\\
& 0.3252 & & 2.3308 & & 0.4462 &\\
$A$& 0.2509 & (0.0855, 0.4467) & 0.6892 & (0.4826, 0.8335) & 0.3122 & (0.1314, 0.5100)\\
& 0.2454 & & 0.6998 & & 0.3085 & \\
\hline
\end{tabular}
\end{footnotesize}
\end{table}

\begin{figure}[H]
\begin{center}
\caption{Densities of the elements of the one-step-ahead  covariance matrices for Normal and Student data estimated using  (a) DPM and Normal and  (b) DPM and Student errors.}
\label{Simulation2}
\subcaption{}
\includegraphics[width=1\textwidth]{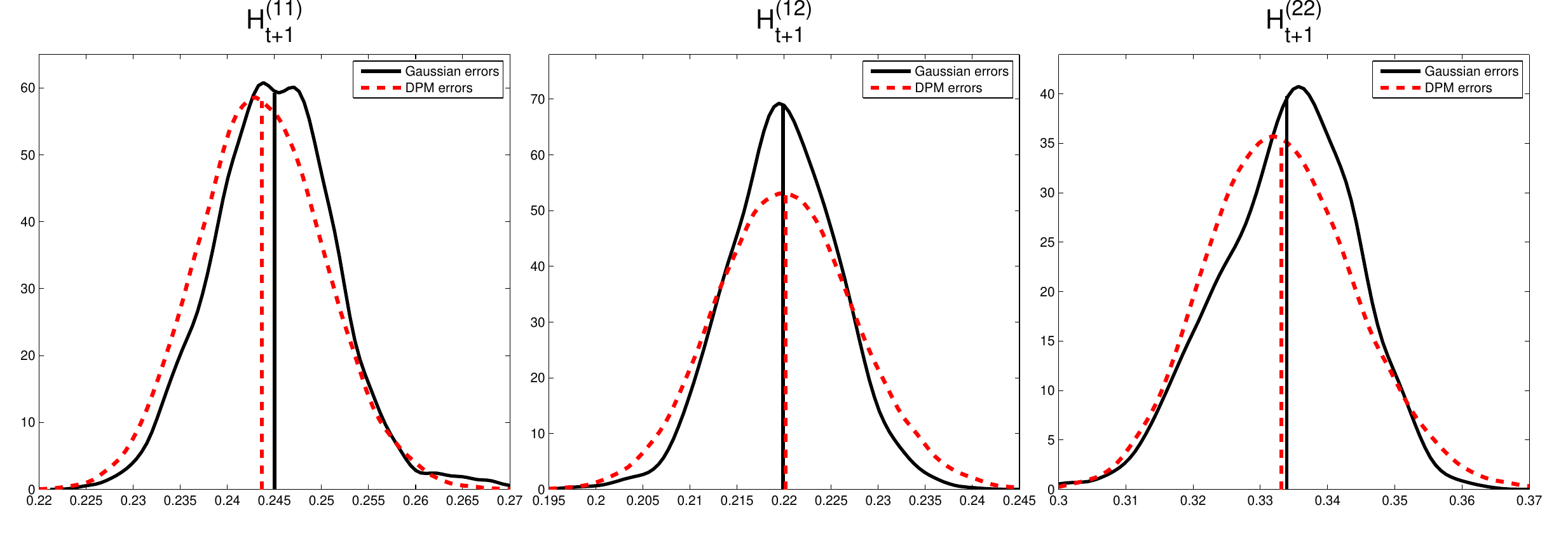}
\subcaption{}
\includegraphics[width=1\textwidth]{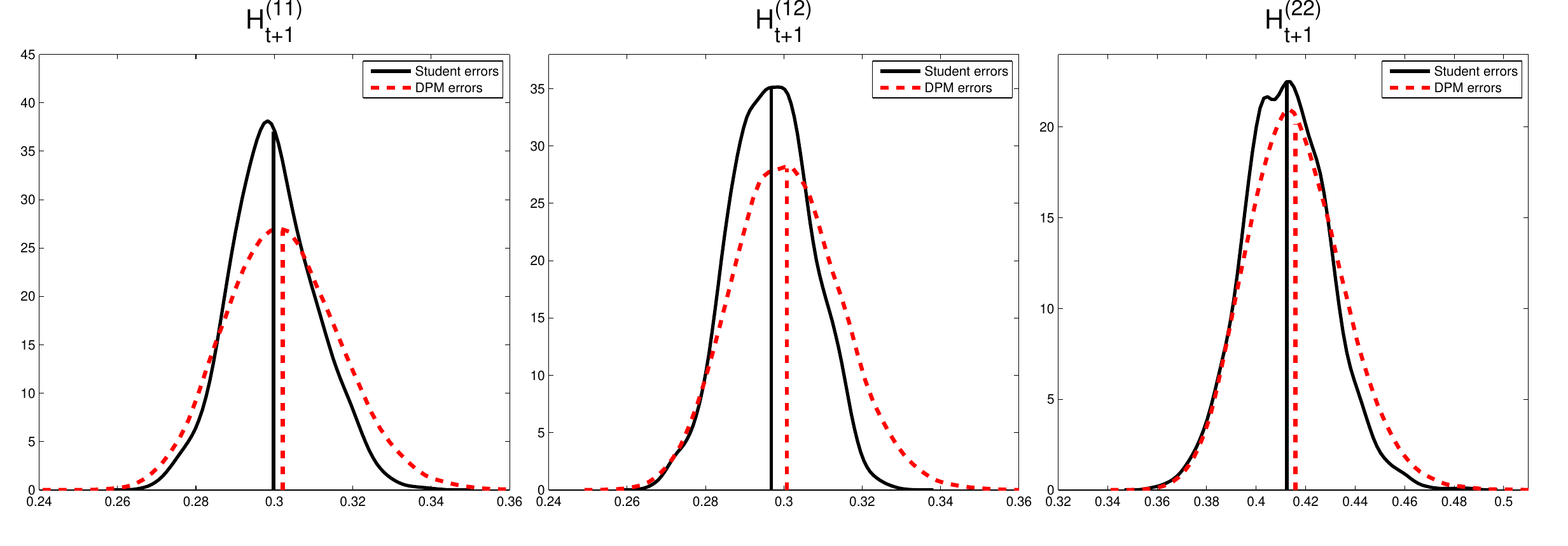}
\end{center}
\end{figure}

Finally, we have estimated the generated Normal and Student data sets assuming Gaussian and Student's t-distributions, respectively. We used the RWMH with 10,000 burn-in plus 40,000 iterations. This way we were able to obtain a sample of 
one-step-ahead covariance matrices $\{H_{t+1}^{(m)}\}_{m=1}^M$, estimated using the true return distributions. Figure \ref{Simulation2} compares the  densities for one-step-ahead covariances $\{H_{t+1}^{\star(m)}\}_{m=1}^M$ assuming a DPM for (a) and (b) with the true data generating model: Gaussian and Student-t respectively. As we can see, the mean estimates and the width and shape of the posterior distributions are very similar for DPM and the ones obtained using the  true return distribution. Therefore, we can conclude that DPM model can adjust to different frequently used distributions for the return data without making any restrictive distributional assumptions.

\section{REAL DATA AND RESULTS}\label{S:data_results}

In this section, we illustrate the performance of the proposed methodology using a real dataset and solve a portfolio allocation problem as described in Section~\ref{S:Applications}.

\subsection{Estimation}

\begin{figure}[H]
\begin{center}
\caption{Log-returns and histograms of Apple Inc.\ and NASDAQ Ind.\ Index.}
\label{returns2}
\includegraphics[width=1\textwidth]{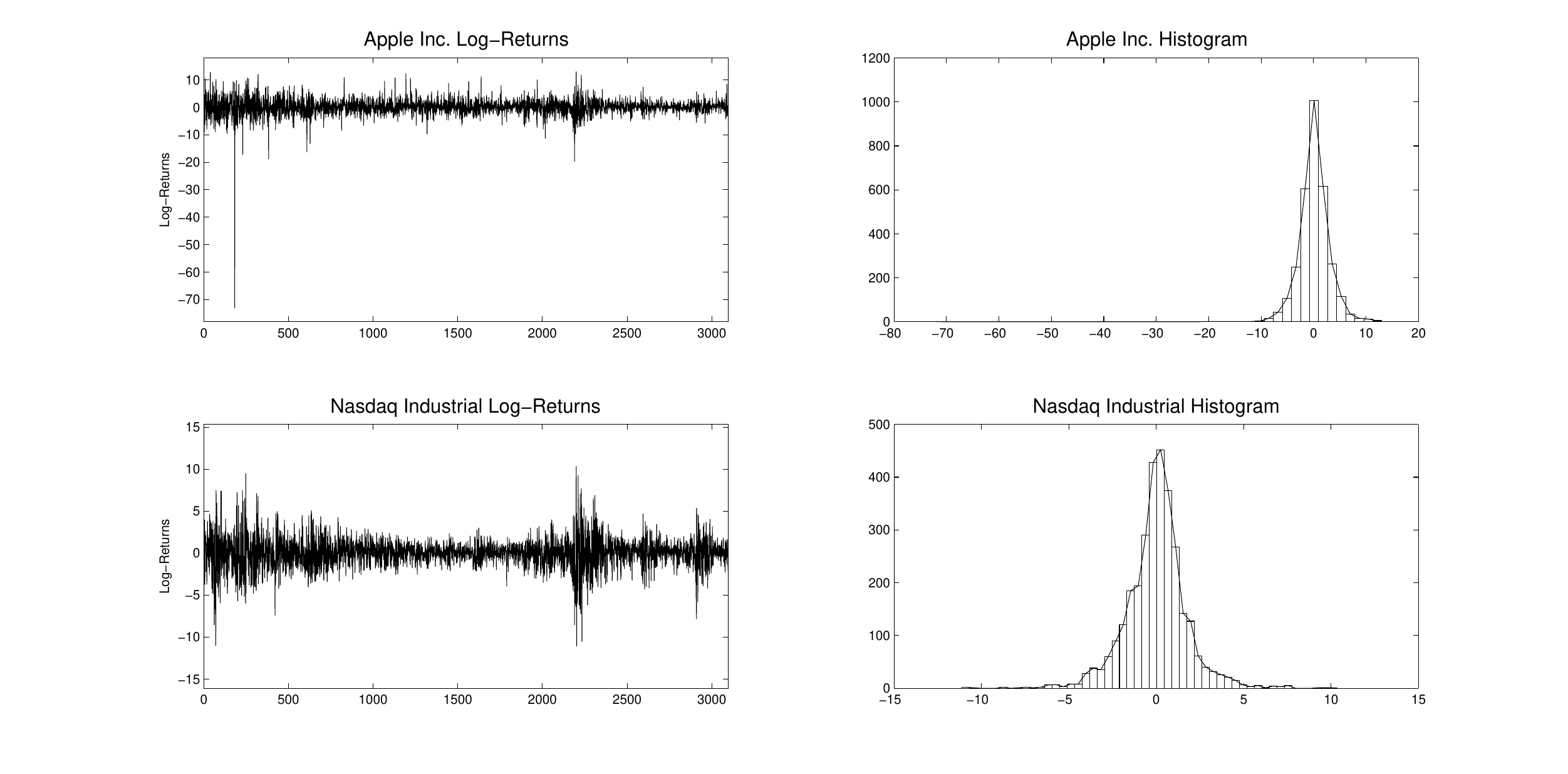}
\end{center}
\end{figure}

We consider the daily price data of Apple Inc.\ company ($P_t^A$) and NASDAQ Industrial index ($P_t^N$) from January 1, 2000 till May 7, 2012, obtained from Yahoo Finance. Then, daily prices are transformed into daily logarithmic returns (in \%), resulting in $T=3098$ observations. Table \ref{table:descriptive_table} provides the basic descriptive statistics and Figure \ref{returns2} illustrates the dynamics of returns.

\begin{table}[tbp]
\centering
\caption{Descriptive statistics of the Apple Inc.\ and NASDAQ Ind.\ return series.}
\label{table:descriptive_table}
\begin{footnotesize}
\begin{tabular}{ccc}
\hline\hline
& $100\times \ln \left(\frac{P_t^A}{P_{t-1}^A}\right)$&$100\times \ln \left(\frac{P_t^N}{P_{t-1}^N}\right)$\\
\hline
Mean &$0.0973$ & $0.0020$\\
Median &$0.1007$ & $0.0766$\\
Variance &$9.7482$ & $3.1537$\\
Skewness &$-4.2492$ & $-0.1487$\\
Kurtosis &$102.0411$ & $7.1513$\\
Correlation &$0.5376$\\
\hline
\end{tabular}
\end{footnotesize}
\end{table}

As expected, the Apple Inc.\ has higher overall variance because of the higher mean return. Both returns do not exhibit any evidence of autoregressive behavior. Apple Inc.\ returns contain one atypical data point, corresponding to September 29, 2000. The very low return is due to an announcement the day before about lower than expected sales.

\begin{figure}[H]
\begin{center}
\caption{Histogram of the posterior sample of $A=c/(1+c)$.}
\label{c_hist}
\includegraphics[width=1\textwidth]{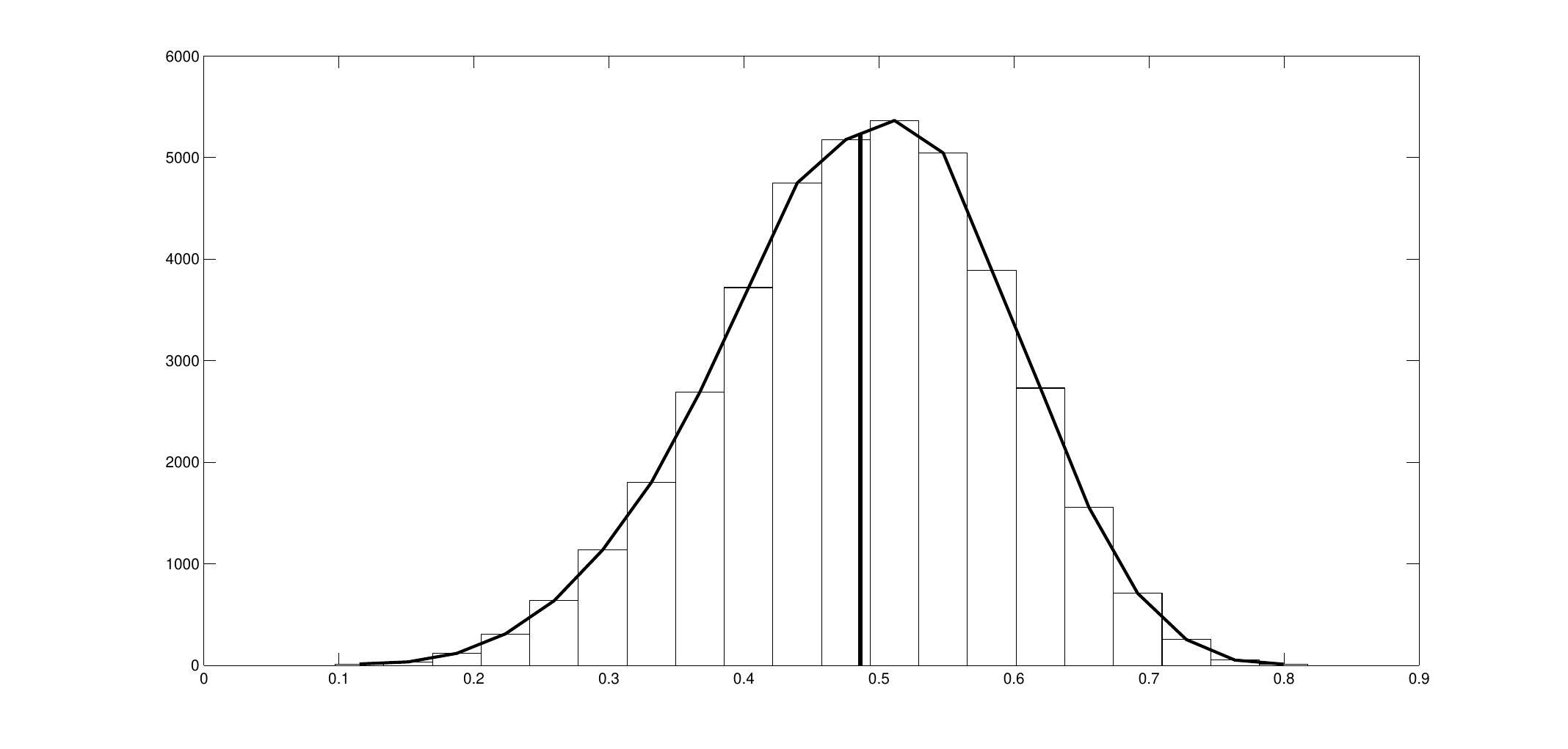}
\end{center}
\end{figure}

Next, the return series was estimated assuming Gaussian, Student-t and DPM errors. Table \ref{table:est_apple} reports the parameter estimation results for the ADCC model assuming a Gaussian, a Student-t and the proposed DPM model for the innovation distribution. As we can see from the table, the constant volatility parameter, $\omega_1$, for the first series is  overestimated under the Gaussian assumption. This is because the Gaussian model does not allow for fat tails, and therefore, all the volatility is summed into the $\omega_1$. Same happens with the asymmetric volatility parameters $\phi_1$ and $\phi_2$ under the Gaussian assumption. The posterior mean of the degrees of freedom under the Student-t is around $7$. Under the DPM model, the average number of non-empty clusters is $z^{\star}=7.8$. Finally, the posterior mean of $A$ is rather close to 0.5, which suggest the better adequacy of the DPM model when compared with the Student-t specification. This is more clear in Figure \ref{c_hist}, which draws the histogram of the posterior distribution for $A$. Note that the posterior probability that $A$ is larger than 0.7 is very small. These results can be compared to the ones obtained in simulation study: when the data comes from a Student-t distribution, DPM estimates a large number of clusters (around 19 in our case) and parameter $A$ is closer to 0.7. On the other hand, when the underlying distribution is Gaussian, DPM model estimates few clusters and parameter $A$ close to 0.2. The results indicate that  for this specific data set neither Gaussian, nor Student-t distributions are appropriate, since the data comes from a distribution, which is positioned in between. 

\begin{table}[tbp]
\caption{Estimation results for Apple Inc.\ (1) and NASDAQ Ind.\ (2) returns asumming a Gaussian, a Student-t and the proposed DPM model for the innovation distribution.}
\centering
\label{table:est_apple}
\begin{footnotesize}
\begin{tabular}{c c c c c c c}
\hline\hline
  \toprule
  & \multicolumn{2}{c}{Gaussian}& \multicolumn{2}{c}{Student-t} & \multicolumn{2}{c}{DPM}  \\

&Mean & $95\%$ CI &Mean & $95\%$ CI&Mean & $95\%$ CI\\
\\\hline
$\omega_1$ 	& 0.2653 &  (0.1603, 0.3942)  &  0.1071  & (0.0619, 0.1659) & 0.1999  & (0.1330, 0.2364) \\
$\omega_2$ 	& 0.0285  & (0.0203, 0.0395)  & 0.0192  & (0.0125, 0.0269) &  0.0088   & (0.0052,  0.0131) \\
$\alpha_1$		& 0.0894  & (0.0651, 0.1244)  & 0.0403 & (0.0279, 0.0534) &  0.0747 &  (0.0575,    0.0964) \\
$\alpha_2$		&  0.0126 &  (0.0020, 0.0270) & 0.0109  & (0.0012, 0.0245) & 0.0055  & (0.0006,    0.0123) \\
$\beta_1$		& 0.8430  & (0.8039,0.8740)  & 0.8975 & (0.8730, 0.9204)  & 0.8771   &  (0.8489,    0.9000) \\
$\beta_2$		& 0.9237 & (0.9052, 0.9375)  & 0.9281  & (0.9121, 0.9408) &  0.9232 & (0.9074,    0.9362) \\
$\phi_1$			& 0.1197 &  (0.0622, 0.1627)  & 0.0409  & (0.0183, 0.0683) &  0.0539    &  (0.0139, 0.0984) \\
$\phi_2$			& 0.1050  & (0.0796, 0.1312)  & 0.0739  & (0.0520, 0.0912) &  0.0370     &  (0.0255, 0.0476)\\
$\kappa$			& 0.0093  & (0.0030, 0.0266)  & 0.0075 & (0.0020, 0.0153) &  0.0172     &  (0.0043, 0.0326) \\
$\lambda$		& 0.9828  & (0.9415, 0.9936)  & 0.9711  & (0.9477, 0.9858) & 0.8914     & (0.8445, 0.9409)\\
$\delta$			& 0.0061 & (0.0008, 0.0213)   & 0.0220  & (0.0094, 0.0392)& 0.0242&  (0.0024,  0.0426) \\
$\nu$					&				  &  									   & 7.1879  & (7.0529, 7.2862)& &  \\
$z^{\star}$				   &				  &  										& 				&									 & 7.7481 &(4.0000, 13.0000) \\
$A$						&  			  &  									   & 					 & 								 &  0.4862 & (0.2723, 0.6746) \\
\hline
\end{tabular}
\end{footnotesize}
\end{table}
   
\begin{figure}[H]
\begin{center}
\caption{Contours of the predictive densities for $r_{T+1}$ under a Gaussian, Student-t and DPM specification for the innovation distribution.}
\label{figure:contours}
\includegraphics[width=1\textwidth]{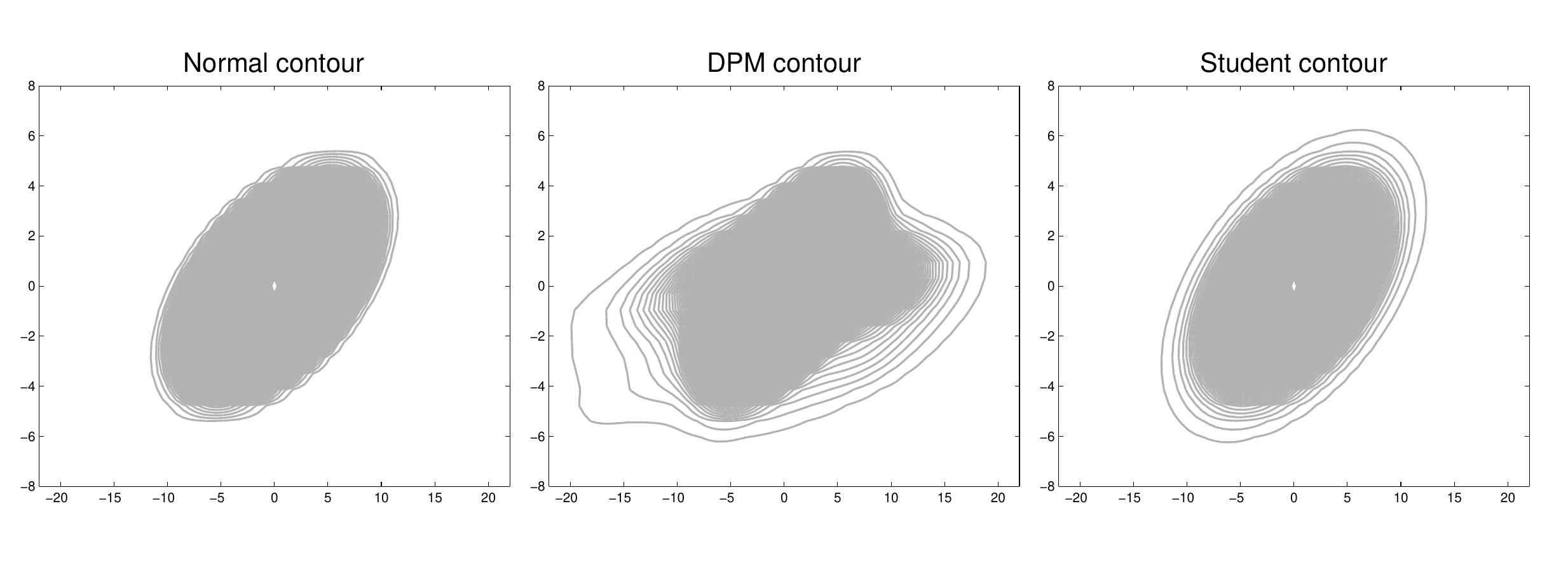}
\end{center}
\end{figure}

\begin{figure}[H]
\begin{center}
\caption{Marginal log-predictive densities for the one-step-ahead Apple (left) and NASDAQ (right) return data, under a Gaussian, Student-t and DPM specification for the innovation distribution.}
\label{figure:log_predictive}
\includegraphics[width=1\textwidth]{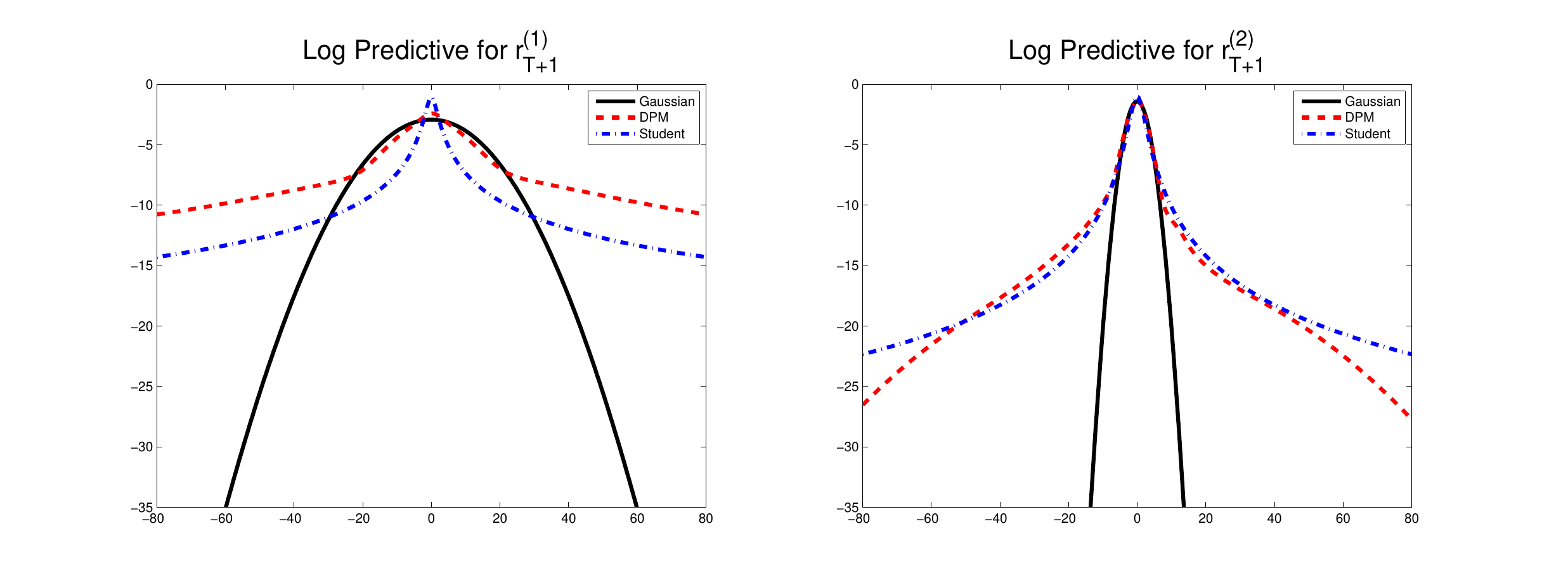}
\end{center}
\end{figure}

Figure \ref{figure:contours} compares the predictive densities of the one-step-ahead return, $r_{T+1}$, under the Gaussian, the Student-t and the proposed DPM specifications for the innovation distribution. Observe that both Normal and Student-t models lead to symmetric predictive densities, although for the Student-t case it exhibits fatter tails, which are completely defined by a single parameter, $\nu$. On the contrary, the  DPM model, which allows for different variances and non zero means, predicts an asymmetric multimodal density with fatter tails. Also, Figure \ref{figure:log_predictive} presents the marginal log predictive densities for the returns under the three specifications. Observe that the DPM model can differentiate between volatile and not so volatile returns, since it predicts obviously fatter tails for the Apple return data, meanwhile for not so volatile NASDAQ data, the difference between DPM and Student-t is not so big. The Gaussian specification in both cases cannot capture the high kurtosis.

Next, following \cite{Jensen2013}, we compare the three estimated models using predictive likelihoods based on a small set of out-of-sample observations,  $\left\{{T+1,\hdots,T+k}\right\}$. For each new observation, we calculate the predictive likelihood as an average over the MCMC iterations given by:
\begin{equation*}
p(r_{T+i}|r^{T+i-1}) = \frac{1}{M}\sum\limits_{m=1}^M p(r_{T+i}|r_{T+i-1},\Theta^{(m)}),\text{ for } i=1,\ldots,k,
\end{equation*}
and then calculate the sum of the logarithms over the out-of-sample time period:
\begin{equation*}
\log p(r_{T+1},\ldots,r_{T+k}) = \sum\limits_{i=1}^k \log p(r_{T+i}|r^{T+i-1}).
\end{equation*}
However, note that, differently to \cite{Jensen2013}, we do not re-estimate the model whenever a new observation arrives to avoid a high increase in the computational cost, we use the already estimated model parameters up to time $T$. Table \ref{table:log_pred} presents the cumulative log-predictive likelihood for the three models using $k=233$ out-of-sample observations. Observe that the sum of the log-predictive favors the DPM model. The difference between the two competing models is the log of a Bayes factor.

\begin{table}[tbp]
\caption{Cumulative log-predictive likelihoods for the DPM, Student and Gaussian error models.}
\centering
\label{table:log_pred}
\begin{footnotesize}
\begin{tabular}{c c}
\hline\hline
Model & $\log p(r_{T+1},\hdots,r_{T+k}) $\\
\hline
DPM & -746.5467\\
Student &-749.2850 \\
Normal & -775.8299\\
\# of out-of-sample obs. & $k=233$\\
\hline
\end{tabular}
\end{footnotesize}
\end{table}

\begin{figure}[H]
\begin{center}
\caption{Posterior Distributions of One-step-ahead Volatilities for the Three Errors Specifications}
\label{fig:omegas}
\includegraphics[width=1\textwidth]{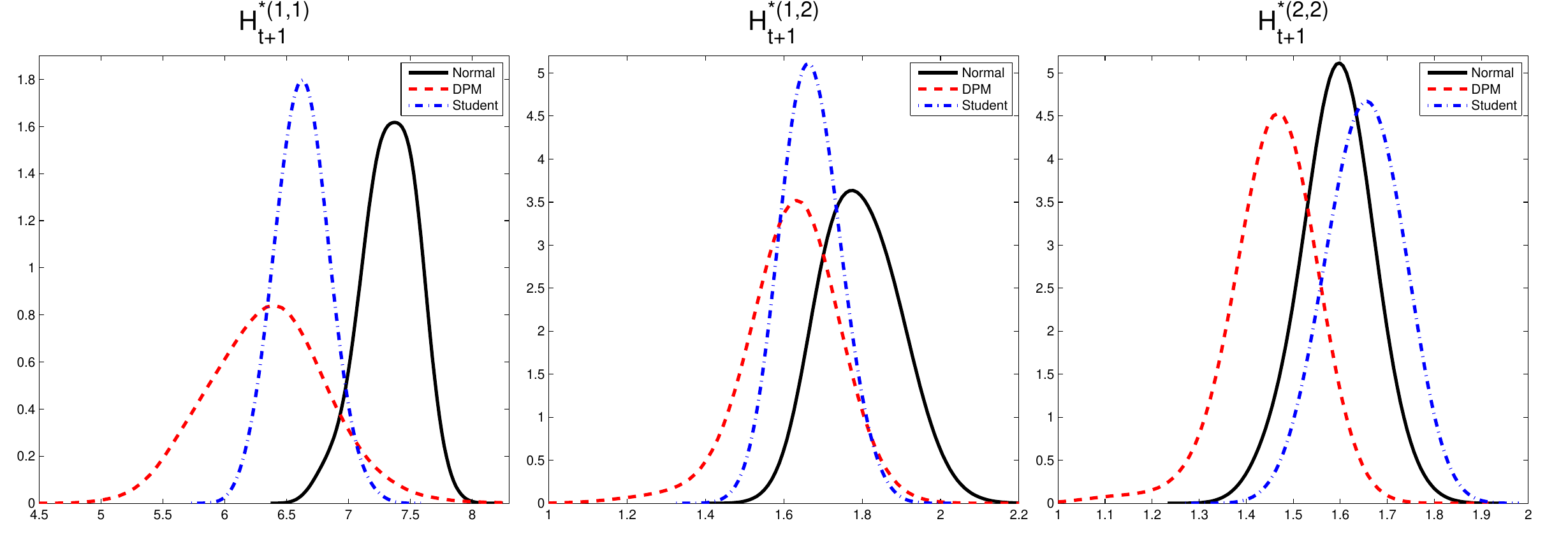}
\end{center}
\end{figure}

 Figure \ref{fig:omegas} draws the posterior densities of the volatilities and Table \ref{table:volatilities} presents the posterior means, medians and confidence intervals for the elements of the one-step-ahead volatility matrix under the Gaussian, the Student-t and the proposed DPM specification. For the latter, these are obtained using (\ref{Eadvolat}) and the explanations given in Section 2.3. Observe that the credible intervals for the DPM model are wider, especially for the marginal one-step-ahead volatility of the first series. This is because it allows for some very volatile mixture components, due to the atypical data point, which seems to provide a more realistic evaluation of risk for an agent. Also, under the DPM model, the posterior distributions for the volatilities are not symmetric.  Finally, Table \ref{table:volatilities} also shows the estimation results for the degrees of freedom parameter under the Student-t assumption, which indicates heavy tails, and for the one-step-ahead mean under the DPM model using (\ref{Eadmean}).

\begin{table}[tbp]
\caption{Posterior means, medians and confidence intervals for the elements of the one-step-ahead volatility matrix under the Gaussian, Student-t and DPM specification.}
\centering
\label{table:volatilities}
\begin{footnotesize}
\begin{tabular}{c c c c c c c}
\hline\hline
  \toprule
  & \multicolumn{2}{c}{Gaussian}& \multicolumn{2}{c}{Student-t} & \multicolumn{2}{c}{DPM}  \\
&Mean & $95\%$ CI &Mean & $95\%$ CI&Mean & $95\%$ CI\\
&Median&&Median & &Median & \\
\\\hline
$H_{T+1}^{{\star}(1,1)}$ 	&  7.3315 &  (6.8461, 7.6932)  &  6.6228      & (6.2307,    7.0214) &  6.3519 & (5.4437,  7.3185) \\
& 7.3419 && 6.6237  & &  6.3532& \\
$H_{T+1}^{{\star}(1,2)}$ 	& 1.7954 &  (1.6379, 1.9831)  &  1.6684       & (1.5609,  1.7944 ) & 1.6215 & (1.3615, 1.8238) \\
&  1.7890&& 1.6649 &  & 1.6284&  \\
$H_{T+1}^{{\star}(2,2)}$ 	& 1.5932 &  (1.4534, 1.7163)  & 1.6524         & (1.5188,    1.7733) & 1.4570 & (1.2624, 1.5953) \\
&  1.5954&&1.6541  &  &1.4634 &  \\
$\nu$ 	&  &   &   7.1879 & (7.0529, 7.2865) &  &  \\
&  &  & 7.1933 &  & &  \\
$\mu_{T+1}^{{\star}(1)}$ 	& 0 &   & 0   &  & 0.1503 & (0.0547,  0.2456) \\
&  &  & &  &0.1503 &  \\
$\mu_{T+1}^{{\star}(2)}$ 	&  0 &  &  0 &  &0.0131 & (-0.0311, 0.0574) \\
&  &  & &  &  0.0132 &  \\
\hline
\end{tabular}
\end{footnotesize}
\end{table}

\subsection{Portfolio Allocation}

\begin{figure}[H]
\begin{center}
\caption{Mean Cumsum of the Width of the 95\% Credible Intervals for Portfolio Weights and Variance.}
\label{fig:p_cis}
\includegraphics[width=1\textwidth]{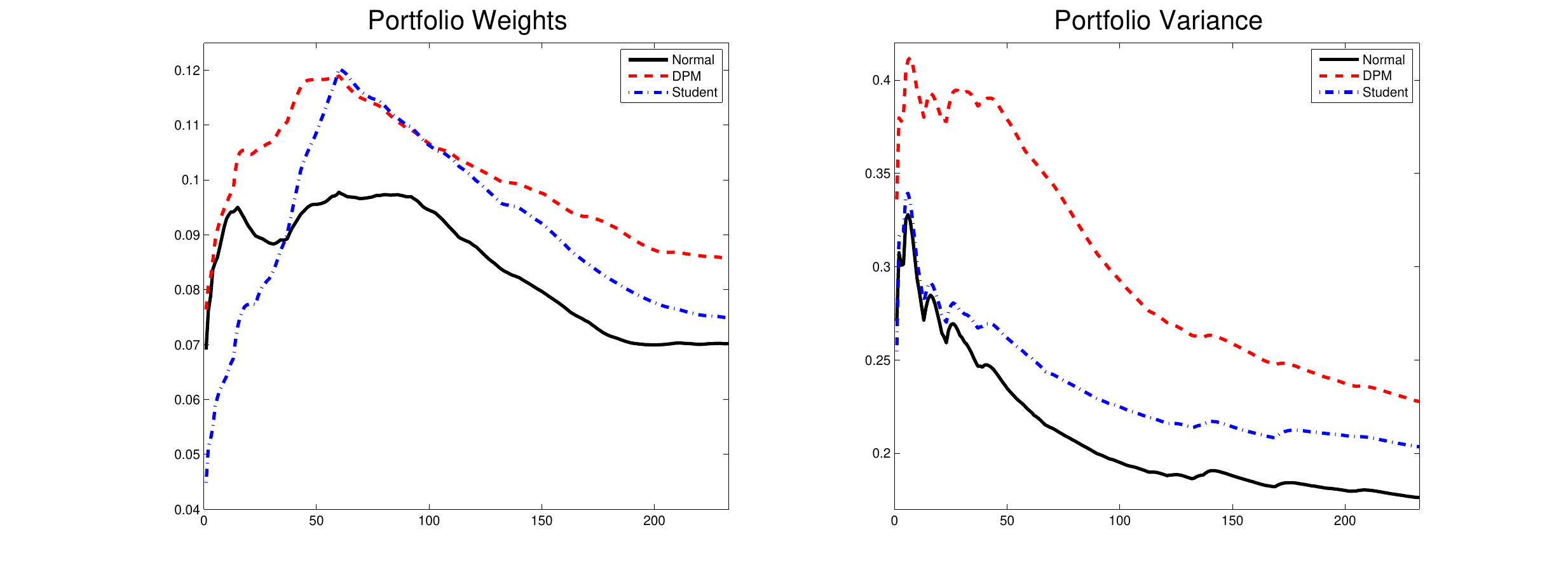}
\end{center}
\end{figure}

Here we are interested in estimating the GMV optimal portfolio of the two real assets, without the short-sale constraint, using the procedure described in Section \ref{S:Applications}. Firstly, we will make predictions on the optimal one-step-ahead portfolio and then, we will consider all the $233$ out-of-sample future observations, adjusting the optimal portfolio weights at each time period. The estimation results for the $T+1$ period are presented in Table \ref{table:portfolio}. The major difference between the estimates is the width of the credible intervals of the portfolio variance. The same can be observed for the rest of the period, as seen in Figure \ref{fig:p_cis}. Therefore, if the investor chooses to be Gaussian or Student, she would underestimate the variance of the variance of her portfolio.

\begin{table}[tbp]
\caption{Posterior mean, median and $95\%$ credible intervals for the optimal one-step-ahead portfolio weight, variance and return.}
\centering
\label{table:portfolio}
\begin{footnotesize}
\begin{tabular}{c c c c c c c}
\hline\hline
  \toprule
  & \multicolumn{2}{c}{Gaussian}& \multicolumn{2}{c}{Student-t} & \multicolumn{2}{c}{DPM}  \\
&Mean & $95\%$ CI &Mean & $95\%$ CI&Mean & $95\%$ CI\\
&Median& &Median &&Median &\\
\\\hline
$p_{T+1}^{\star}$ 	& -0.0401  &  ( -0.0775,   -0.0084)  &  -0.0015    & (-0.0245,    0.0204 ) &  -0.0327 & (-0.0726, 0.0038) \\
&  -0.0391 &  &  -0.0012 &  &  -0.0320 &  \\
$\sigma_{T+1}^{2{\star}}$		& 1.5811     & (1.4609,    1.7318)  &1.6518    & (1.5175,    1.7722) & 1.4526    &  (1.2577, 1.5939) \\
&  1.5771&  &1.6536& & 1.4591 & \\
$g_{T+1}^{\star}$		& 3.2009     & (3.1255,  3.2897)  & 3.1091     & (3.0572,  3.1637)  & 3.1832    &  (3.0965,    3.2780 ) \\
	& 3.1986 &  & 3.1085 & &   3.1817    &  \\
\hline
\end{tabular}
\end{footnotesize}
\end{table}
           
\begin{figure}[H]
\begin{center}
\caption{Dynamics of Portfolio Weights and 95\% Credible Intervals.}
\label{fig:p_weights}
\includegraphics[width=1\textwidth]{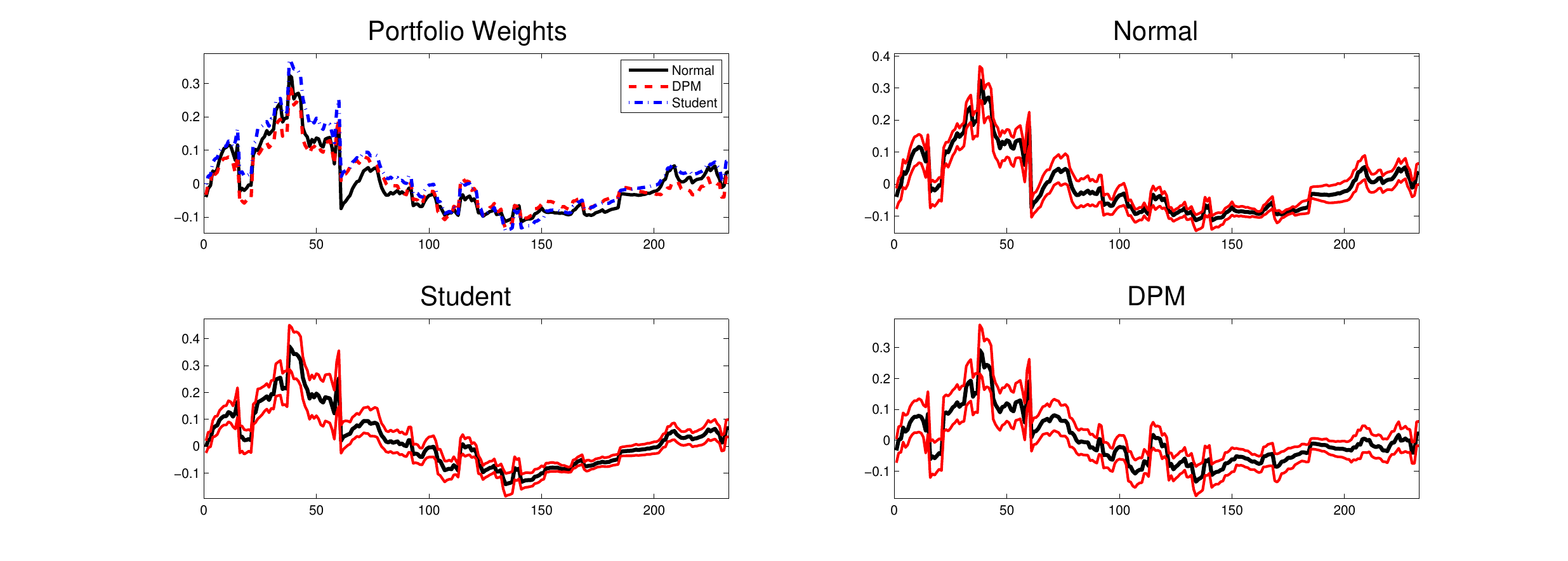}
\end{center}
\end{figure}

\begin{figure}[H]
\begin{center}
\caption{Dynamics of Portfolio Variance and 95\% Credible Intervals.}
\label{fig:p_variance}
\includegraphics[width=1\textwidth]{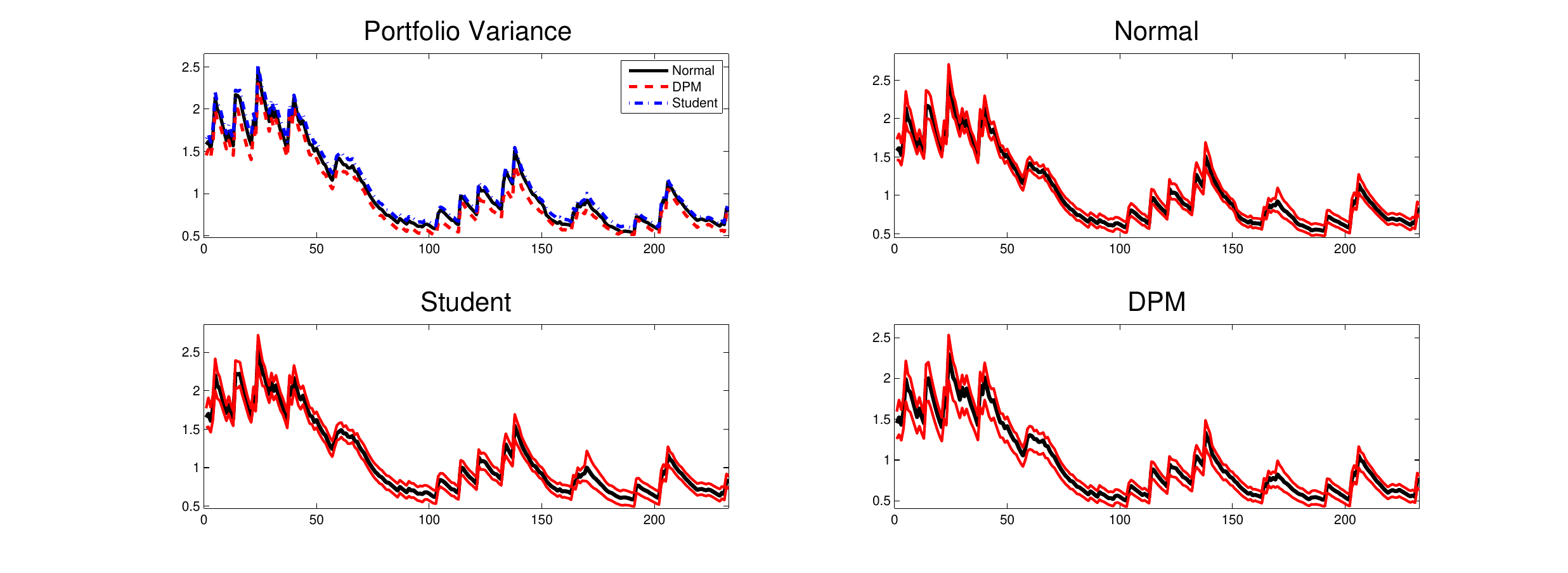}
\end{center}
\end{figure}

Next, we estimate the optimal portfolio weights for the entire out-of-sample period of $233$ observations. Figures \ref{fig:p_weights} and \ref{fig:p_variance} present the dynamics of the estimated portfolio weights and variances for each of the models. It shows that along time the mean portfolio weights are rather similar across all three models. As mentioned before, the differences arise in the thickness of the credible intervals for the portfolio variance, as seen in Figure \ref{fig:p_cis}.  This allows for a more realistic evaluation of the uncertainty that investor is facing in financial risk management problems.

To sum up, these portfolio allocation exercises helped to illustrate the direct consequences of return distribution to the uncertainty of financial decisions. The DPM model permits the investor to perform inference and prediction about the returns and their volatilities without imposing arbitrary restrictions on the data generating process. In the portfolio allocation context, adjusting portfolio weights at each period might lead to high transaction costs, thus the investor will adjust her portfolio only if the expected utility after the adjustment minus the transaction costs is greater than the expected utility without the adjustment. The illustration has shown the differences in error specifications in using real data. We have illustrated how quantification of uncertainty reflects distributional assumptions of the errors.

\section{Conclusions}\label{S:Conclusion}

In this paper we have proposed a Bayesian non-parametric approach for modeling the distribution of multiple returns. We have used an ADCC-GJR-GARCH model to explain the individual volatilities and the time-varying correlations and taking into consideration the asymmetries in individual assets' volatilities, as well as in the correlations. The errors are modeled using a location-scale mixture of infinite Gaussian distributions that has been shown to allow for a great flexibility in the return distribution in terms of skewness and kurtosis. An MCMC method has been presented for model estimation and prediction. For that, a Dirichlet process mixture (DPM) prior has been given
to the infinite mixture of multivariate Gaussian distribution. We have also considered a dynamic portfolio allocation problem, where the time-varying covariance matrix is estimated using a ADCC-GJR-GARCH model. We have presented a short simulation study that illustrates the differences arising from different assumptions for the errors and shows the adaptability of the DPM model. The simulation results suggest that the proposed approach appears to be able to fit adequately several frequently used distributions. Here is the main importance of our approach because in practice one never knows which is the true error distribution. Finally, we have presented an application to
return data of Apple Inc.\ and NASDAQ Industrial that compares the DPM specification with a Gaussian and Student-t distributions. Model comparison via log-predictive likelihood favors the non-parametric approach. Additionally,
we have employed the proposed approach to solve a portfolio allocation problem. In the application we have showed that even though the point estimates for optimal portfolio weights are very similar for Gaussian or Student-t, the non-parametric credible intervals  for the volatilities are wider. Therefore, the normality or Student assumptions forces the investor to be overconfident about her estimates. The explained methodology and obtained results are not limited to this specific risk management problem and could be expanded into various other topics in applied finance and risk management.

\section*{Acknowledgements}

The first and second authors are grateful for the financial support from MEC grant ECO2011-25706. The third author acknowledges financial support from MEC grant ECO2012-38442.

\bibliographystyle{model5-names}

\end{document}